\documentclass[conference]{IEEEtran}
\IEEEoverridecommandlockouts
\usepackage{amssymb,amsmath,bm}
\usepackage{amsfonts}
\usepackage{cite}
\usepackage{graphicx,subfigure}
\usepackage{psfrag}
\usepackage{url}
\usepackage[latin1]{inputenc}
\usepackage[absolute,overlay]{textpos}
\usepackage{tikz}
\usetikzlibrary{arrows,calc,decorations.markings}
\usetikzlibrary{topaths}
\usetikzlibrary{arrows}
\usetikzlibrary{shadows}
\usetikzlibrary{positioning}
\usetikzlibrary{matrix}
\usetikzlibrary{shapes.geometric}
\usetikzlibrary{decorations.pathmorphing}
\usetikzlibrary{calc}
\usetikzlibrary{fit}					
\usetikzlibrary{backgrounds}	

\usepackage{pgf}
\usetikzlibrary{arrows,automata}
\usepackage[latin1]{inputenc}
\usepackage[nocomma]{optidef}

\usepackage{algorithmic}
\usepackage{graphicx}
\usepackage{textcomp}
\usepackage{xcolor}
\def\BibTeX{{\rm B\kern-.05em{\sc i\kern-.025em b}\kern-.08em
    T\kern-.1667em\lower.7ex\hbox{E}\kern-.125emX}}
    
\usepackage{multicol}
\usepackage{algorithm}
\usepackage{algorithmic}

\makeatletter
\def\thickhline{%
  \noalign{\ifnum0=`}\fi\hrule \@height \thickarrayrulewidth \futurelet
   \reserved@a\@xthickhline}
\def\@xthickhline{\ifx\reserved@a\thickhline
               \vskip\doublerulesep
               \vskip-\thickarrayrulewidth
             \fi
      \ifnum0=`{\fi}}
\makeatother

\newlength{\thickarrayrulewidth}
\usepackage{threeparttable}
\usepackage{booktabs}

\usepackage{arydshln}

\makeatletter
\newcommand{\linebreakand}{%
  \end{@IEEEauthorhalign}
  \hfill\mbox{}\par
  \mbox{}\hfill\begin{@IEEEauthorhalign}
}
\makeatother

\begin{document}

\title{Massive Wireless Energy Transfer with Multiple Power Beacons for very large Internet of Things}

\author{
    \IEEEauthorblockN{Osmel Mart\'{i}nez Rosabal\IEEEauthorrefmark{1}, Onel L. Alcaraz L\'{o}pez\IEEEauthorrefmark{1}, Hirley Alves\IEEEauthorrefmark{1}, Richard D. Souza\IEEEauthorrefmark{2}, Samuel Montejo-S\'{a}nchez\IEEEauthorrefmark{3}}
    
    \IEEEauthorblockA{
    \IEEEauthorrefmark{1} Centre for Wireless Communications (CWC), Oulu, Finland \\ 
    \IEEEauthorrefmark{2} Federal University of Santa Catarina (UFSC), Florian\' {o}polis, Brazil \\ 
    \IEEEauthorrefmark{3} Programa Institucional de Fomento a la I+D+i, Universidad Tecnol\'{o}gica  Metropolitana, Santiago, Chile \\
    }
    \{osmel.martinezrosabal, onel.alcarazlopez, hirley.alves\}@oulu.fi, richard.demo@ufsc.br, smontejo@utem.cl
}
\maketitle

\begin{abstract}
The Internet of Things (IoT) comprises an increasing number of low-power and low-cost devices that autonomously interact with the surrounding environment. As a consequence of their popularity, future IoT deployments will be massive, which demands energy-efficient systems to extend their lifetime and improve the user experience. Radio frequency wireless energy transfer has the potential of powering massive IoT networks, thus eliminating the need for frequent battery replacement by using the so-called power beacons (PBs). In this paper, we provide a framework for minimizing the sum transmit power of the PBs using devices' positions information and their current battery state. Our strategy aims to reduce the PBs' power consumption and to mitigate the possible impact of the electromagnetic radiation on human health. We also present analytical insights for the case of very distant clusters and evaluate their applicability. Numerical results show that our proposed framework reduces the outage probability as the number of PBs and/or the energy demands increase.
\end{abstract}

\begin{IEEEkeywords}
clustering, massive IoT, optimal PBs deployment, RF-WET, energy outage
\end{IEEEkeywords}

\section{Introduction}
The recent advances in wireless communications technologies have brought a plethora of new opportunities for developing the so-called Internet of Things (IoT). In short, IoT defines a network of uniquely identified physical objects that autonomously sense, actuate, compute and exchange information throughout the Internet \cite{younan2020challenges}. In general, ``an object" can be anything from a smart ring, electronic appliances, or sensors that feedback environmental measurements. Currently, this paradigm has taken over in many spheres of our daily life and society such as in industrial automation, smart healthcare, supply chain, smart infrastructure, and social \& business applications \cite{goyal2021internet}. Not surprisingly, the number of connected devices is constantly growing, and by 2030 the estimations target massive deployments of 71 billion IoT devices \cite{McKinseyReport2020}. 

One of the big challenges of such applications is extending their lifetime as IoT devices are generally powered by tiny batteries. Besides, regular maintenance becomes impractical or too costly in most cases due to the massive number of devices or the hard-to-reach installation conditions, such as in medical implants, waste management, and civil infrastructure monitoring applications \cite{9075807}. This issue can cause service interruptions due to power outages in the network, which degrades the end-user experience and hence lowers the profits of the networks' operators \cite{zeadally2020design}. In multi-hop IoT applications, for instance, power outages in few devices can affect the network's performance, as some of them have the role of forwarding the packets of the far away devices to the central receiver \cite{8343216}.

To extend the lifetime of IoT applications, radio-frequency (RF) wireless energy transfer (WET) envisions a network where the so-called power beacons (PBs) wirelessly replenish devices' batteries or even sustain the operation of batteryless devices by means of electromagnetic waves \cite{mahmood2020white}. Moreover, RF-WET exhibits other desirable features such as energy broadcast to power multiple users simultaneously, service provisioning under non-line-of-sight conditions, and the potential to charge moving IoT devices \cite{7867826, 9319211}. By making RF-WET a ubiquitous service, one can deal with the high heterogeneity of IoT networks and guarantee different levels of quality-of-service (QoS) in terms of energy demands. Hence, as RF-WET reduces frequent maintenance, it could considerably diminish water and air pollution caused by toxic chemicals that are released into the environment after inadequate battery recycling \cite{7565191}.

A key component in RF energy harvesting (EH) systems is the rectenna, which comprises \cite{8951078}: i) receiving antenna(s); ii) a matching network; iii) a rectifier circuit; and iv) a low-pass filter. At the receiver, the rectenna turns the microwave energy into a direct current (DC) used in most of the circuits of IoT devices. Compared with traditional wireless information decoders, rectennas require a significant amount of incident RF power to provide a non-zero DC output, typically in the order of $-10$~dBm \cite{7081084}. However, the channel impairments such as distant-dependent loss, large- and small- scale fading processes cannot be often overcome to deliver such incident power levels by deliberately increasing the transmit power. In this direction, energy beamforming algorithms offer a promising solution as the intention is to steer sharp energy beams directly to the intended devices avoiding spreading the energy where there is no device \cite{8269301}. However, realizing efficient energy beamforming may be challenging in practical RF-WET systems. The reasons are:
\begin{itemize}
    \item On one hand, accurate CSI is required, which is difficult/costly to acquire, and its benefits vanish, in massive deployments \cite{9319211}. For this reason, some initiatives have been proposed to prescind from instantaneous CSI at the price of transmitting with suboptimal beamforming strategies, e.g., \cite{9184149, 6884811}, using statistical CSI and received energy feedback, respectively. It is worth noticing, that although both are sub-optimal strategies, their performance becomes asymptotically optimal in the presence of strong line-of-sight components and large channel training periods, respectively.
    \item On the other hand, energy beamforming at high transmit power increases the RF electromagnetic field radiation (RF-EMF) in relatively small incident areas whose risk on human health has concerned the research community \cite{9013993, 8245790, 8485951}. To evaluate the impact of the RF signals on the human health, the most frequently used metrics are: the specific absorption rate (SAR) and the plane-wave equivalent power density (PD) \cite{9020454}.
\end{itemize}

Hence, the transmission schemes for enabling massive WET can be designed instead, over the basis of side-information such as devices' positions, statistical CSI, or even received energy feedback from the IoT devices \cite{lopez2021csi}. At the same time, these methods must ensure sufficient energy to meet stringent QoS requirements while guaranteeing the human safety against high RF-EMF levels exposure. Last, but not less important, the algorithms for efficiently deploying the PBs also play a key role to eliminate blind spots in the network, and to distribute the energy according to the specific requirements.

\subsection{Related work}
The research community has put effort into the design of PBs' deployment algorithms to meet stringent QoS requirements in the coming IoT use cases, e.g., \cite{ding2020optimal, 9209664, 8428435, 8714083, 9310236}. The authors in \cite{ding2020optimal} proposed a PBs' deployment algorithm that guarantees a network-wide energy availability at the devices while considering the installation/maintenance costs of PBs. They conducted experiments to show that: i) the complexity of proposed algorithms decreases as the network densifies since each PB charges more devices simultaneously; ii) PBs installation cost gets more expensive as the per-device minimum energy requirements increase; and iii) when using directional antennas at the PBs, the deployment cost reduces as the beamwidth of the antennas slim down at the price of reducing the incident RF power in the network. In \cite{9209664}, the authors considered the scenario where a mobile PB charges the service area while moving along the area perimeter to minimize the charging time. Therein, the authors divided the service area into smaller sub-partitions and focused on the instantaneous worst position, which changes as the PB moves. In \cite{8428435}, the authors studied the PBs' placement problem aimed to maximize the overall harvested energy. They proposed an algorithm for optimizing the positions and orientations of PBs using a piecewise linear EH model, and evaluated its performance using both numerical simulations and field experiments. In addition, \cite{8714083} proposed an algorithm for finding the positions and orientation of a set of PBs with different hardware capabilities to maximize the overall harvested energy. Finally, the optimal deployment of PBs for charging a massive IoT network was studied in \cite{9310236}. Therein, the authors optimized the PBs' positions to maximize the foreseen minimum average RF energy available in the area, since neither instantaneous CSI nor devices' locations information was available. Based on simulation results, the authors promoted distributed deployments of PBs over the installation of a centred PB radiating with the same total power to overcome the dominant impact of the distance-dependent loss. 

Moreover, other works have also considered safety metrics in the WET optimization. For instance, in \cite{9013993} the authors studied a smart healthcare application where PBs charge wearable devices. Therein, they minimize the maximum incident power while providing sufficient energy for the wearables. In \cite{8245790}, the authors proposed a framework for optimizing the position and height of a distributed antenna system, where each antenna is connected via underground lines to a central PB, considering a safe radiation level. In \cite{8485951}, it is maximized the network-wide harvested energy such that the probability of exceeding a certain RF-EMF is met \cite{9020454}.

It is worth noticing that none of these works exploits simultaneously sensors' position and battery state information. Hence, their performance has limited capabilities to act proactively to face the imminent ``death" of one of its IoT devices after the total discharge of its battery.

\subsection{Contributions}
Different from previous works, herein we aim to minimize the sum transmit power of the PBs such that the energy available at devices' batteries suffices for proper operation. We exploit sensors' position information to divide the IoT network into clusters each headed by a PB; and the current battery state to proactively optimize the power allocation strategy. 

Our main contributions are summarized below.
\begin{itemize}
    \item We propose a framework for optimizing PBs' positions and power allocation based on the devices' positions and their current battery state;
    \item We provide analytical approximations to the power allocation problem that holds when the contribution of the head PBs dominates the incident RF power within their corresponding cluster;
    \item Our optimization strategy has two main implications: i) to reduce the energy consumption at the PBs, possibly powered by a limited source, and ii) mitigate the possible impact of the EMF radiation on human health;
    \item Our results show that the outage probability decreases as the number of deployed PBs increases and/or the energy demands grow with respect to the minimum battery level for proper operation.
\end{itemize}
  
\subsection{Organization of the paper}
Next, Section \ref{section:system_model} introduces the system model, and Section \ref{section:problem_formulation} presents the problem formulation. We discuss the methods for finding the optimal PBs' positions and computing the power allocation in Section \ref{section:optimization_framework}, while in Section \ref{section:practical_cosiderations} we discuss some practical considerations for implementing our strategy. In Section \ref{section:numerical_results}, we show and analyze numerical results, and finally, Section \ref{section:conclusions} concludes the paper.

\textit{Notation:} Herein, we use boldface lowercase letters to denote column vectors, e.g., $\mathbf{x} = \{x_i\}$. The operator $|\cdot|$ can represent either the absolute value for scalars or the cardinality of a set, while $\lVert x \lVert_p = (\sum_{\forall i \ge 1} \vert x^p_i \rvert)^{1/p}$ denotes the $\mathcal{\ell}_p$-norm \cite[eq. (1)]{lange2013derivatives}. Moreover, $\mathbb{P}[A]$ is the probability of event $A$ and $\mathbb{E}[x]$ denotes the expectation of the randon variable $x$, whereas $\mathcal{O}(\cdot)$ is the big-O notation, which specifies worst-case complexity. Table~\ref{tab:notation} lists the symbols used throughout the paper.
\begin{table}[t]
    \centering
    \caption{List of symbols}
    \label{tab:notation}
    \setlength{\tabcolsep}{10pt}
    \resizebox{\columnwidth}{!}{\begin{tabular}{p{1.2cm} p{7cm}}
        \thickhline
            \textbf{Parameter} & \textbf{Definition} \\
        \hline
            $\mathrm{S}_j$ & $j^\text{th}$ IoT device in the set $\mathcal{S}$ \\
            $\mathrm{PB}_i$ & $i^\text{th}$ PB in the set $\mathcal{B}$ \\
            $\varrho_{j,i}$ & path-loss function of the $\mathrm{PB}_i \rightarrow \mathrm{EH}_j$ link\\
            $d_{j,i}$ & distance of the $\mathrm{PB}_i \rightarrow \mathrm{EH}_j$ link\\
            $\{p_i\}$ & transmit power of $\mathrm{PB}_i$ \\
            $p_\mathrm{max}$ & PBs' maximum transmit power \\
            $x_{i}$ & energy-carrying signal transmitted from $\mathrm{PB}_i$ \\
            $\tau$ & duration of a charging time slot\\ 
            $t$ & time slot index \\
            $\{\mathbf{u_j}\}$ & IoT devices' locations \\
            $\{\mathbf{u_i}\}$ & PBs' positions \\
            $E_j^{(t)}$ & battery level of $S_j$ in the time slot $t$\\
            $\alpha_j^{(t)}$ & activation state of $S_j$ in the time slot $t$\\
            $\Delta E_j^{(t)}$ & energy consumed by $S_j$ during the time slot $t$\\
            $\xi_j^{(t)}$ & harvested energy of $S_j$ during the time slot $t$\\
            $\tilde{\xi}_j^{(t)}$ & estimated $\xi_j^{(t)}$ for the time slot $t$\\
            $\mathcal{G}, \mathcal{G}^{-1}$ & EH function and its inverse respectively \\
            $G_i$ & antenna gain of $\mathrm{PB}_i$ \\
            $G_j$ & antenna gain of $S_j$ \\
            $E_\mathrm{th}$ & target energy level at the batteries \\
            $E_\mathrm{max}$ & devices' battery capacity \\
            $P_\mathrm{active}$ & devices' power consumption in active mode \\
            $P_\mathrm{sleep}$ & devices' power consumption in sleep mode \\
        \thickhline
    \end{tabular}}
\end{table} 
\section{System model}\label{section:system_model}
\begin{figure}[t!]
	\centering
	\includegraphics[width=\columnwidth]{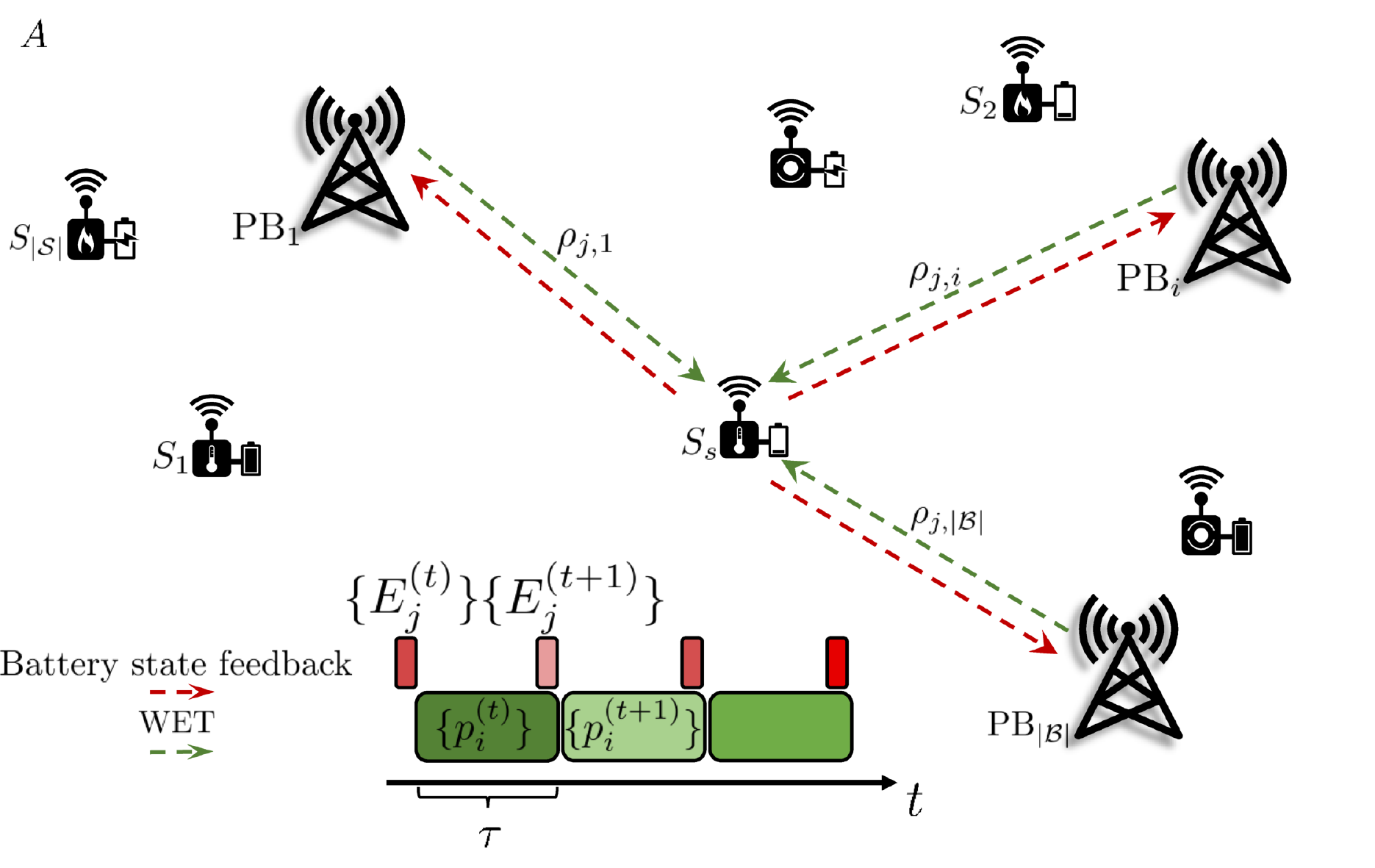}
	\caption{System model.}
	\vspace*{-2mm}		
	\label{fig:system_model}
\end{figure}
Let us consider that a set $\mathcal{B} = \{\mathrm{PB}_i|i=1,2,\ldots,|\mathcal{B}|\}$ of PBs wirelessly powers a set $\mathcal{S} = \{S_j|j=1,2,\ldots,|\mathcal{S}|\}$ of IoT devices deployed in a certain area $A$ as shown in Figure~\ref{fig:system_model}. Each $\mathrm{PB}_i$ is equipped with an omnidirectional antenna and transmits an energy-carrying signal $x_i$ that satisfies $\mathbb{E}[|x_i x_{i'}|] = 0$ with $i \ne i'$, and $\mathbb{E}[|x_i|^2] = p_i$, $\forall \mathrm{PB}_i, \mathrm{PB}_{i'} \in \mathcal{B}$, where $p_i$ denotes the transmit power. We assume that the positions of the IoT devices are known\footnote{In many massive IoT applications, devices tend to have fixed positions or limited mobility, therefore position knowledge can be acquired at deployment or estimated via previous transmissions \cite{9269936}.} and denoted by $\mathbf{u}_j \in \mathbb{R}^2$, while PBs locations  have coordinates $\mathbf{u}_i \in \mathbb{R}^2$, both in the Cartesian system. The distance of the link $\mathrm{PB}_i \rightarrow S_j$ is denoted by $d_{j,i} = \lVert \mathbf{u}_j - \mathbf{u}_i \rVert_2$, while its path loss is given by
\begin{equation}
    \rho_{j,i} = \mathcal{F} (d_{j,i}).
\end{equation}
Each IoT device $S_j$ carries a rechargeable battery and harvests
\begin{equation}\label{eq:EH}
    \xi_j^{(t)} = \tau \mathcal{G}\left(\sum_{i=1}^{|\mathcal{S}|} p_i^{(t)} \rho_{j,i}\right),
\end{equation}
energy units in a time slot $t$ of duration $\tau$. Herein, we consider that each charging time slot is sufficiently long so channel fluctuations may approximately average out, thus, impacting little the performance. The function $\mathcal{G}:\mathbb{R}^+ \rightarrow \mathbb{R}^+$ characterizes the non-linear energy harvesting model of the IoT devices and distinguishes three regions \cite{7547357}: i) zero-output power, in which the incident RF power is below the sensitivity of the EH circuit; ii) input power-dependent region, where the output DC power is an increasing non-linear function of the input RF power; iii) saturation region, in which the output DC power becomes constant and independent of the input signal.

Finally, the residual battery energy of device $S_j$ is denoted by $E_j$, and its current activation state by $\alpha_j \in \{0, 1\}$. In idle/sleep mode ($\alpha_j = 0$), the power consumption of IoT devices is $P_\mathrm{sleep}$, whereas in active mode ($\alpha_j = 1$), it is $P_\mathrm{active}$. Hence, at the beginning of the charging time slot $t+1$, the battery level is updated according to
\begin{equation}
    E_j^{(t+1)} \leftarrow \min(E_j^{(t)} + \xi_j^{(t)} - \Delta E_j^{(t)}, E_\mathrm{max}),
\end{equation}
where 
\begin{equation}
    \Delta E_j^{(t)} \!=\! \tau \big[(1 - \alpha_j^{(t)}) P_\mathrm{sleep} + \alpha_j^{(t)}) P_\mathrm{active}\big],
\end{equation}
corresponds to the consumed energy during the time slot $t$, and $E_\mathrm{max}$ is the battery capacity.

Moreover, we consider that the devices feedback their current battery state over reliable channels. This information is used by the PBs to optimize their power allocation strategy.

\section{Problem formulation}\label{section:problem_formulation}
Our main goal is to minimize the sum PBs' transmit power such that the actual energy in the batteries $E_j, \forall S_j \in \mathcal{S}$, is above an energy threshold $E_\mathrm{th}$ at the end of the charging time slot. The optimization problem can be formulated as follows
\begin{subequations}\label{P1}
    \begin{alignat}{2}
    \mathbf{P1:} &\underset{\{p_i^{(t)}\}, \{\mathbf{u}_i\}}{\mathrm{min.}} \ && \sum_{b=1}^{B} p_i^{(t)} \label{P1a}\\
    &\quad \text{s.t.} \ && E_j^{(t-1)} + \tilde{\xi}_j^{(t)} \geq E_\mathrm{th}, \ \ \forall S_j \in \mathcal{S}, \label{P1b}\\
    & \ && p_i^{(t)} \leq p_\mathrm{max}, \ \qquad \qquad \forall \mathrm{PB}_i \in \mathcal{B}. \label{P1c}
    \end{alignat}
\end{subequations}
Notice that the constraint \eqref{P1b} considers an estimate harvested energy $\tilde{\xi}_j^{(t)}$ given by \eqref{eq:EH}, that should suffice to replenish devices' batteries above the threshold $E_\mathrm{th}$, during the next time slot. Herein, we consider $\mathcal{F}(\cdot)$ to be known in advance via prior measurement campaigns \cite{9383771} and using machine learning/artificial intelligence \cite{aldossari2019machine} methods, but in practice this will never be exact, and estimation errors will affect somewhat the performance. Moreover, the constraint \eqref{P1c} guarantees that the transmit power doesn't violate PBs' hardware specifications or EMF regulations. 

This problem is non-convex and highly non-linear in general due to the relationship between the distance and the path-loss function and the non-linearity of the function $\mathcal{G}(\cdot)$, and to the best of the authors' knowledge it cannot be written in convex form. Therefore, a global optimum solution is not guaranteed by any solver.

\section{Optimization framework}\label{section:optimization_framework}
To solve \textbf{P1}, we propose a two-step method: i) first optimize PBs' positions by taking advantage of the knowledge about devices' positions while keeping the transmit power constant; and then ii) solve the optimal power allocation problem. 

\subsection{PBs deployment optimization based on clustering}
Herein, we utilize the well-known K-Means clustering algorithm \cite{1056489} to place the PBs as head clusters. Note that since the path-loss function depends on the Euclidean distance between PBs and IoT devices, the K-Means approach seems an appealing choice for solving our problem. However, we propose a slight modification of the original algorithm. Our approach aims to minimize the distance from each PB to the farthest sensor in its cluster, which impacts the ultimate incident power. Thus, the proposed algorithm re-computes the PBs' positions as Chebyshev centres.

Let $\mathcal{S}_i \subset \mathcal{S}$ denotes the subset of IoT devices associated with $\mathrm{PB}_i$ after running the K-Means algorithm, and given by
\begin{equation}\label{eq:kmeans}
    \mathcal{S}_i = \{ \mathbf{u}_j : \lVert \mathbf{u}_j - \mathbf{u}_i \rVert_2 \leq \lVert \mathbf{u}_j - \mathbf{u}_{i'} \rVert_2\},
\end{equation}
with $i' \neq i$, $\forall \mathrm{PB}_i,\mathrm{PB}_{i'} \in \mathcal{B}$. Then, each K-Chebyshev PBs positions is updated according to
\begin{align}
    \mathbf{u}_i = \underset{\mathbf{u} \in \mathbb{R}^2}{\mathrm{argmin}} \ \underset{S_j \in \mathcal{S}_i}{\max} \ \lVert \mathbf{u}_j - \mathbf{u} \rVert_2,
\end{align}
which can be found by solving the following equivalent convex problem
\begin{subequations}\label{P2}
    \begin{alignat}{2}
    \mathbf{P2:} \ &\underset{\mathbf{u}_i, r_i}{\mathrm{min.}} \quad && r_i \label{P2a}\\
    &\text{s.t.} \quad && \lVert \mathbf{u}_j - \mathbf{u}_i \rVert^2 - r_i \leq 0, \quad \forall S_j \in \mathcal{S}_i, \label{P2b}
    \end{alignat}
\end{subequations}
where $\sqrt{r_i}$ is the radius of the minimal-radius circumference enclosing all the devices in $\mathcal{S}_i$. Since \textbf{P2} is now convex, we can solve it using standard solvers packages such as CVX \cite{gb08, cvx}.

Algorithm~\ref{alg1} details the steps for determining the PBs' positioning based on clustering. Notice that the step \ref{alg1:4} determines the cluster each device belongs to, while and step \ref{alg1:5} updates the clusters' centroids according to the mean of each cluster. The loop ends when the centroids' positions do not change significantly, according to the error parameter $\mu$. Finally, once the cluster assignment is done, the algorithm re-computes the clusters' centroids with the solutions of problem \textbf{P2}. 
\begin{algorithm}[t]
\caption{Clustering-based PBs deployment}
\begin{algorithmic}[1]\label{alg1}
\STATE \textbf{Input:} $\{\mathbf{u}_j\}$, $\mu$ \label{alg1:1}
\STATE Set $k = 0$ \label{alg1:2}
\STATE Initialize $\{\mathbf{u}_i^0\}$ \label{alg1:3}
\REPEAT
\STATE Compute $\mathcal{S}_i^{(k)}$ according to \eqref{eq:kmeans} \label{alg1:4}
\STATE $\mathbf{u}_i^{(k+1)} = \frac{1}{|\mathcal{S}_i^{(k)}|} \sum_{\mathbf{u}_j \in \mathcal{S}_i^{k}} \mathbf{u}_j$ \label{alg1:5}
\STATE $k \gets k + 1$
\UNTIL $\lVert \mathbf{u}_i^{(k+1)} - \mathbf{u}_i^{(k)} \rVert_2 \leq \mu$
\STATE Solve \textbf{P2}
\STATE \textbf{Output:} $\{\mathbf{u}_i^{(k+1)}\}$
\end{algorithmic}
\end{algorithm}
\subsection{PBs' power optimization}
Once we find the PBs' positions, \textbf{P1} becomes a linear programming (LP) problem in $\{p_i\}$ since \eqref{P1b} is turned into a linear system in the form
\begin{equation}\label{linear_system}
    \sum_{i=1}^{|\mathcal{B}|} p_i^{(t)} \rho_{j,i} \geq \mathcal{G}^{-1}\left( \frac{E_\mathrm{th} - E_j^{(t-1)}}{\tau} \right), \ \forall S_j \in \mathcal{S},
\end{equation}
where $\mathcal{G}^{-1}(\cdot)$ denotes the inverse of the EH function. Then, 
the resulting LP with $|\mathcal{B}|$ variables and $|\mathcal{S}| + |\mathcal{B}|$ inequality constraints can be efficiently solved using interior-point methods with an accuracy $\epsilon$ in $\mathcal{O}(\sqrt{|\mathcal{S}|}\log{(1/\epsilon)})$ iterations, where each iteration demands $\mathcal{O}(|\mathcal{S}|^3)$ arithmetic operations \cite{nesterov1994interior}.

 \subsubsection{Approximate power allocation strategy}
It is worth noticing that as the clusters are more separated, the contribution of the head PB dominates the incident RF power in its corresponding cluster. This might be the case when the IoT network has a highly sparse deployment of EH IoT devices arranged in very separated clusters, or when the distance-dependent loss is sufficiently strong to neglect the power contribution from neighbour clusters. In both cases, we can neglect the contribution of the neighbour PBs, and the linear system \eqref{linear_system} reduces to
\begin{align}
    p_i^{(t)} \rho_{j,i} &\geq \mathcal{G}^{-1}\left(\frac{E_\mathrm{th} - E_j^{(t-1)}}{\tau}\right), \ \forall S_j \in \mathcal{S}_i, \\
    p_i^{(t)} &\stackrel{(a)}{=}  \underset{j}{\max} \Bigg[\rho^{-1}_{j,i} \mathcal{G}^{-1}\!\left(\! \frac{E_\mathrm{th} \!-\! E_j^{(t-1)}}{\tau}\!\right)\Bigg]\!,  \label{eq:approx}
\end{align}
where $(a)$ comes from considering the device with the higher product of path loss and energy demands. Observe that if $p_i$, computed as in \eqref{eq:approx}, is not greater than $p_\mathrm{max}$, it constitutes a low-complexity solution to the power allocation problem. Otherwise, 
the resulting power allocation is not feasible in the sense that $\mathrm{PB}_i$ is just allowed to transmit with a maximum power $p_\mathrm{max}$, which in this case does not suffice the devices' energy requirements. In any case, the power allocation can be set as
\begin{equation}
    p_i^{*(t)} = \min(p_i^{(t)},p_\mathrm{max}),
\end{equation}
to allow satisfying, at least partially, the devices' energy requirements. Notice that, to obtain the optimal solution $\{p_i^{* (t)}\}$ we just require at most $\mathcal{O}(|\mathcal{S}| + 2|\mathcal{B}|)$ arithmetic operations.

\section{Practical considerations}\label{section:practical_cosiderations}
Herein, we discuss the practicalities for implementing the proposed strategy. First, notice that clustering requires devices' location information, which can be provided during the network planning in static deployments. However, in quasi-static or mobile scenarios, clustering as a positioning strategy may require that also the PBs can move/fly to update their position as the network changes.  

Another detail to consider is that battery state feedback is subject to the channel impairments and hence, not all PBs could reliably receive it all the time. In this direction, we propose that PBs collect devices' battery state and validate this information in a distributed way based on previous updates, statistics of devices' activation probabilities, and devices' power consumption. For instance, the PBs can have a highly reliable link between them, over which they can run a blockchain-based lightweight algorithm to validate all energy transactions \cite{8839968}.

Finally, due to the inherent characteristics of the EH circuit the charging period may take more than one time slot. That is, if the energy requirements of an IoT device surpass the amount it can harvest during the time $\tau$, then the optimization problem will be infeasible and the system will have to schedule additional time slots, until the device finally meets the energy requirements.

\section{Numerical results}\label{section:numerical_results}
In this section, we present numerical results on how to address \textbf{P1}. The default parameters for simulations are listed in Table~\ref{tab:simparam} unless we establish the contrary. 
\begin{table}[t]
    \centering
    \caption{Default simulation parameters.}
    \label{tab:simparam}
    {\begin{tabular}{c c | c c}
         \thickhline
            \textbf{Parameter} & \textbf{Value} & \textbf{Parameter} & \textbf{Value} \\
        \thickhline
            $A$ & $30 \times 15$~m$^2$ & $\tau$ & $120$~s \\
            $E_\mathrm{max}$ & $1$~J & $P_\mathrm{sleep}$ & $10$~$\mu$W \\
            $\alpha_j$ & $\sim \mathrm{Beta}(0.5, 0.5)$ & $P_\mathrm{active}$ & $1$~mW \\
            $f$ & $2.4$~GHz & $G_i G_j$ & $24$ \\
            $S$ & $64$ & $p_\mathrm{max}$ & $4$~W  \\
        \thickhline
    \end{tabular}}
\end{table} 
For the EH circuit of the IoT devices we adopt the sigmoidal-based model in \cite{7264986}
\begin{equation}
    \mathcal{G}(x) = \frac{\varpi (1 - e^{-c_1x})}{(1 + e^{-c_1(x - c_0)})} ,
\end{equation}
whose inverse is
\begin{equation}
    \mathcal{G}^{-1}(y) = -\frac{1}{c_1}\ln{\left(\frac{\varpi - y}{ye^{c_0c_1}+\varpi}\right)},
\end{equation}
where $\varpi = 10.73$~mW is the saturation level, $x$ is the incident RF power, $y$ the harvested energy, and $c_0 = 5.365$, $c_1 = 0.2308$ are unitless constant obtained by standard curve fitting using the measurement data in \cite{4494663}.

As a metric of QoS, we consider the energy outage probability. That is, an IoT device is in outage if the actual battery state at the beginning of the charging slot is insufficient for it to operate for $\tau$ time units. That is,
\begin{equation}
    P_j = \mathbb{P}[E_j^{(t)} < \Delta E_j^{(t)}],
\end{equation}
and the average outage probability, i.e., the average probability of having one device in outage, is
\begin{equation}
    P_\mathrm{out} = \mathbb{E}[P_j].
\end{equation}

Finally, we adopt the log-distance path-loss model 
\begin{equation}
    \rho_{j,i} = G_i G_j\left(\frac{\lambda}{4\pi}\right)^2 d_{j,i}^{-2.7},
\end{equation}
where $G_i$ and $G_j$ denote the antenna gains of  $\mathrm{PB}_i$ and $S_j$, respectively, and $\lambda$ is the wavelength of the energy-carrying signals.

\begin{figure}[t]
	\centering
	\includegraphics[width=\columnwidth]{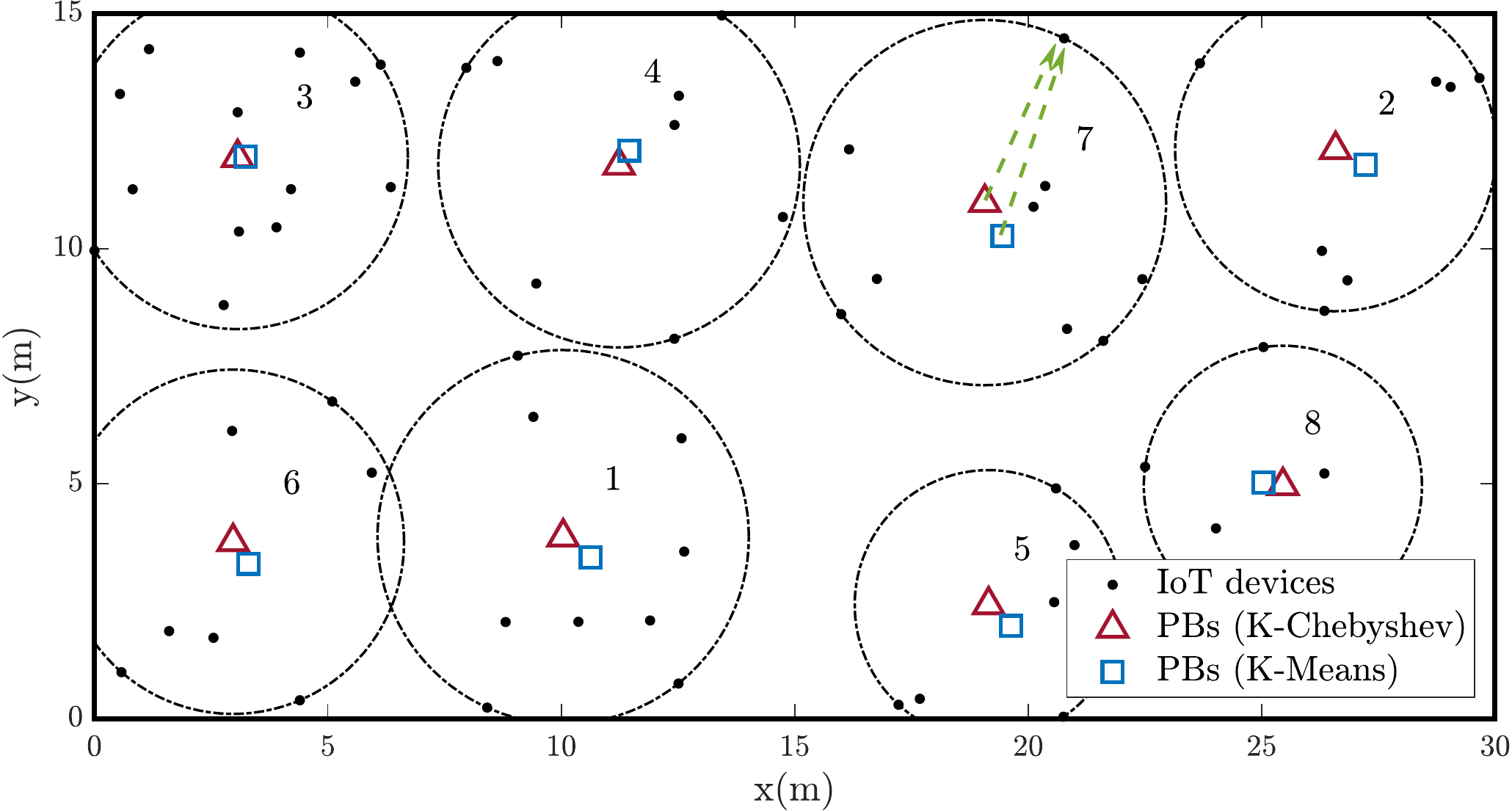}
	\vspace*{-4mm}
	\caption{Clustering-based PBs' positioning approach, where the circumferences represents the clusters' limits and the numbers their corresponding index.}
	\vspace*{-2mm}		
	\label{fig:clustering_powerallocation}
\end{figure}

Figure~\ref{fig:clustering_powerallocation} depicts the clustering strategy for positioning the PBs using both the original K-Means algorithm and the clustering with K-Chebyshev centroids. 
The reader can notice that both strategies lead to different PBs deployments. Indeed, taking the seventh cluster as an example, we can see that the K-Chebyshev centroid has a lower maximum intra-cluster distance with respect to the K-Means solution.

\begin{figure}[t]
	\centering
	\subfigure{\includegraphics[width=\columnwidth]{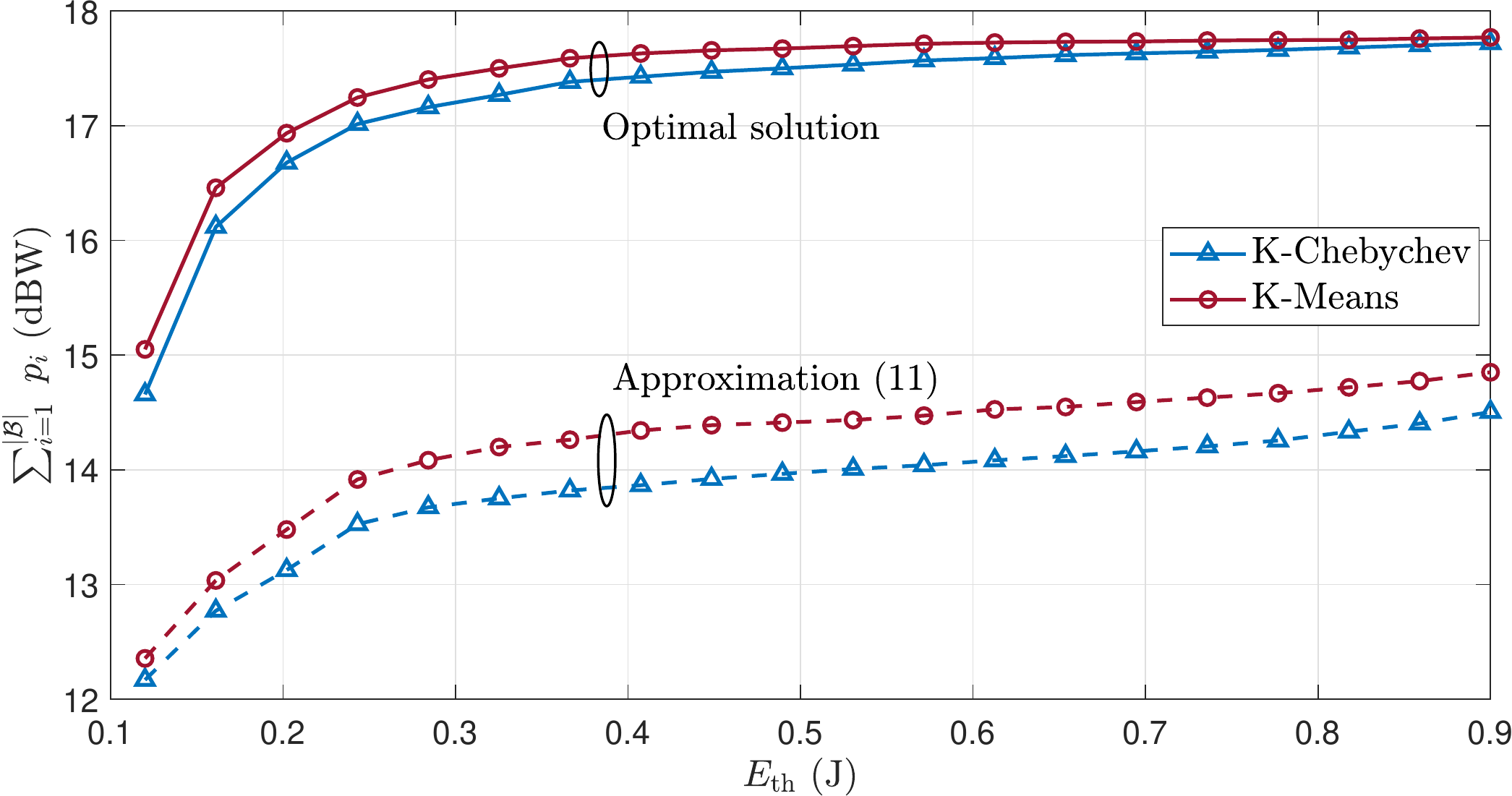}}
	\subfigure{\includegraphics[width=\columnwidth]{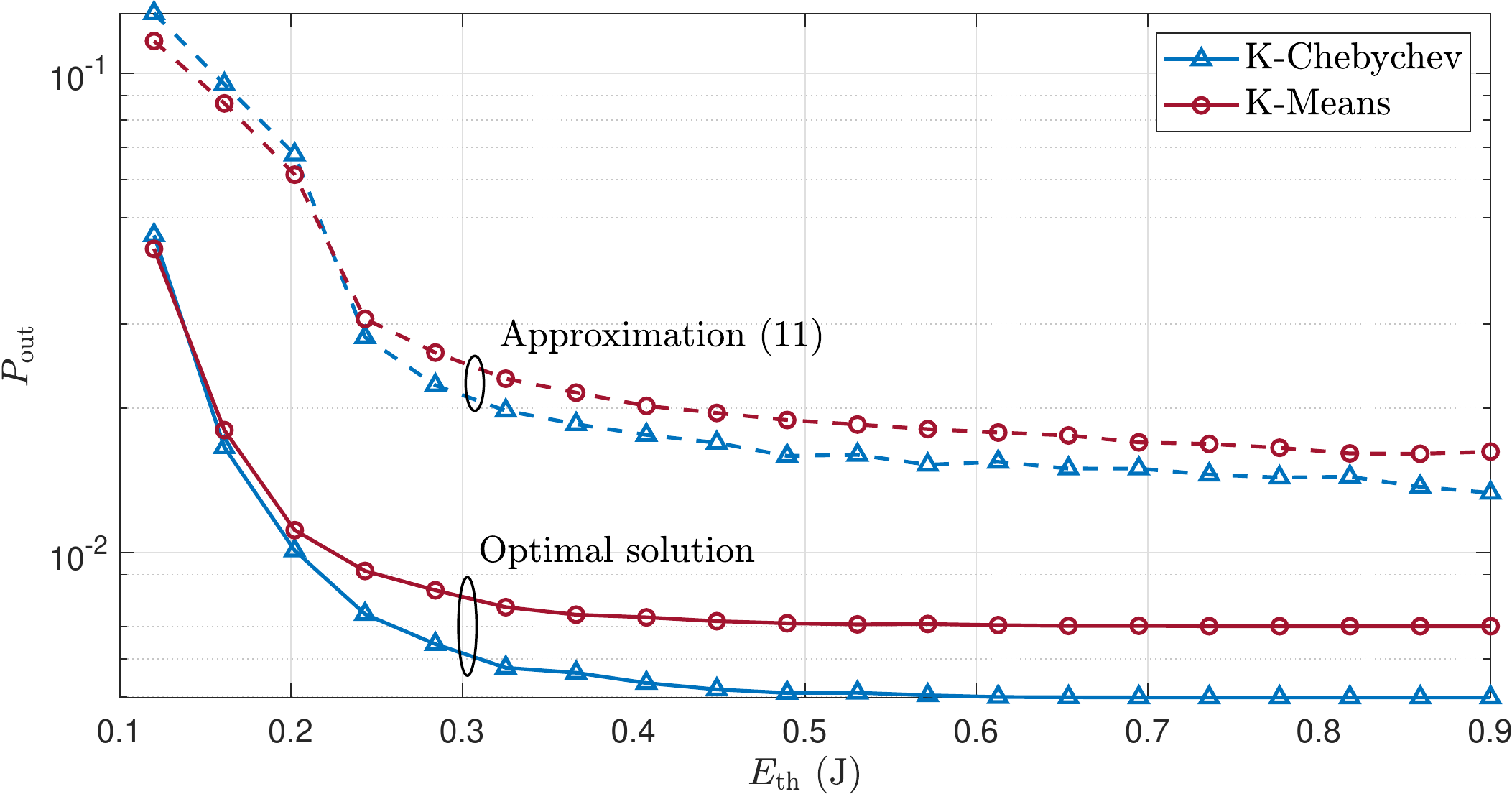}}
	\caption{Network performance vs $E_\mathrm{th}$, for $|\mathcal{B}| = 15$, in terms of: i) sum transmit power (top) and ii) outage probability (bottom).}
	\vspace*{-2mm}		
	\label{fig:vs_E_th}
\end{figure}

Figure~\ref{fig:vs_E_th} illustrates the impact of the energy threshold $E_\mathrm{th}$ on the optimal sum transmit power and the outage probability for $|\mathcal{B}| = 15$.
As the devices' energy demands are tightened, the PBs' transmit power increases and the outage probability decreases. In particular, K-Chebyshev centering achieves better results in almost all cases, as it maximizes the minimum incident RF power within each cluster. Moreover, the approximation~\eqref{eq:approx} achieves lower values of sum transmit power at the cost of a poorer performance in terms of energy outage probability since it considers each cluster as an independent network. From these plots, we can define the optimal energy threshold $E_\mathrm{th}^*$ as the value from which the outage probability doesn't improve significantly. For this particular deployment, we have $E_\mathrm{th}^* = 0.4$~J approximately.

\begin{figure}[t]
	\centering
	\subfigure{\includegraphics[width=\columnwidth]{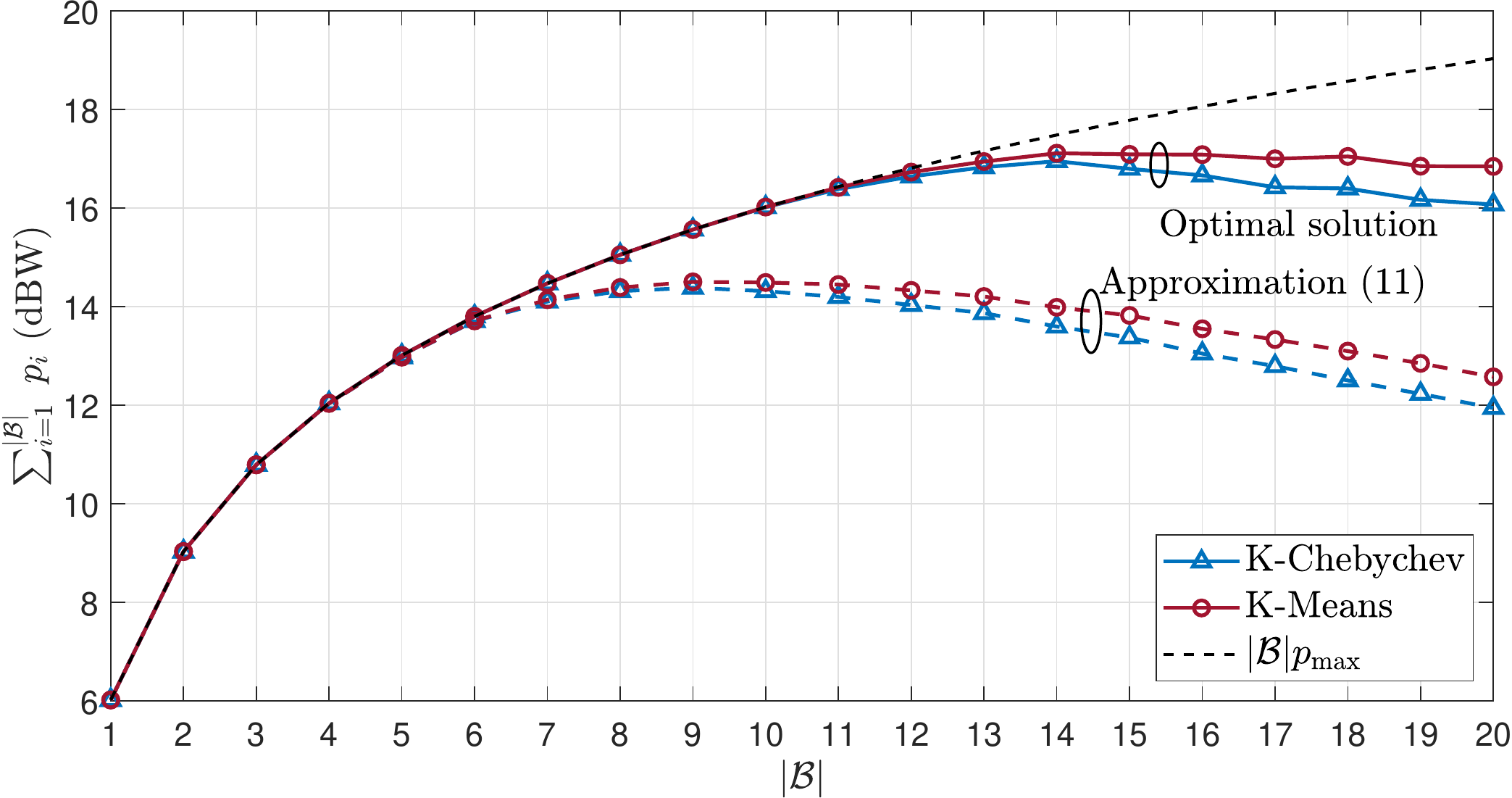}}
	\subfigure{\includegraphics[width=\columnwidth]{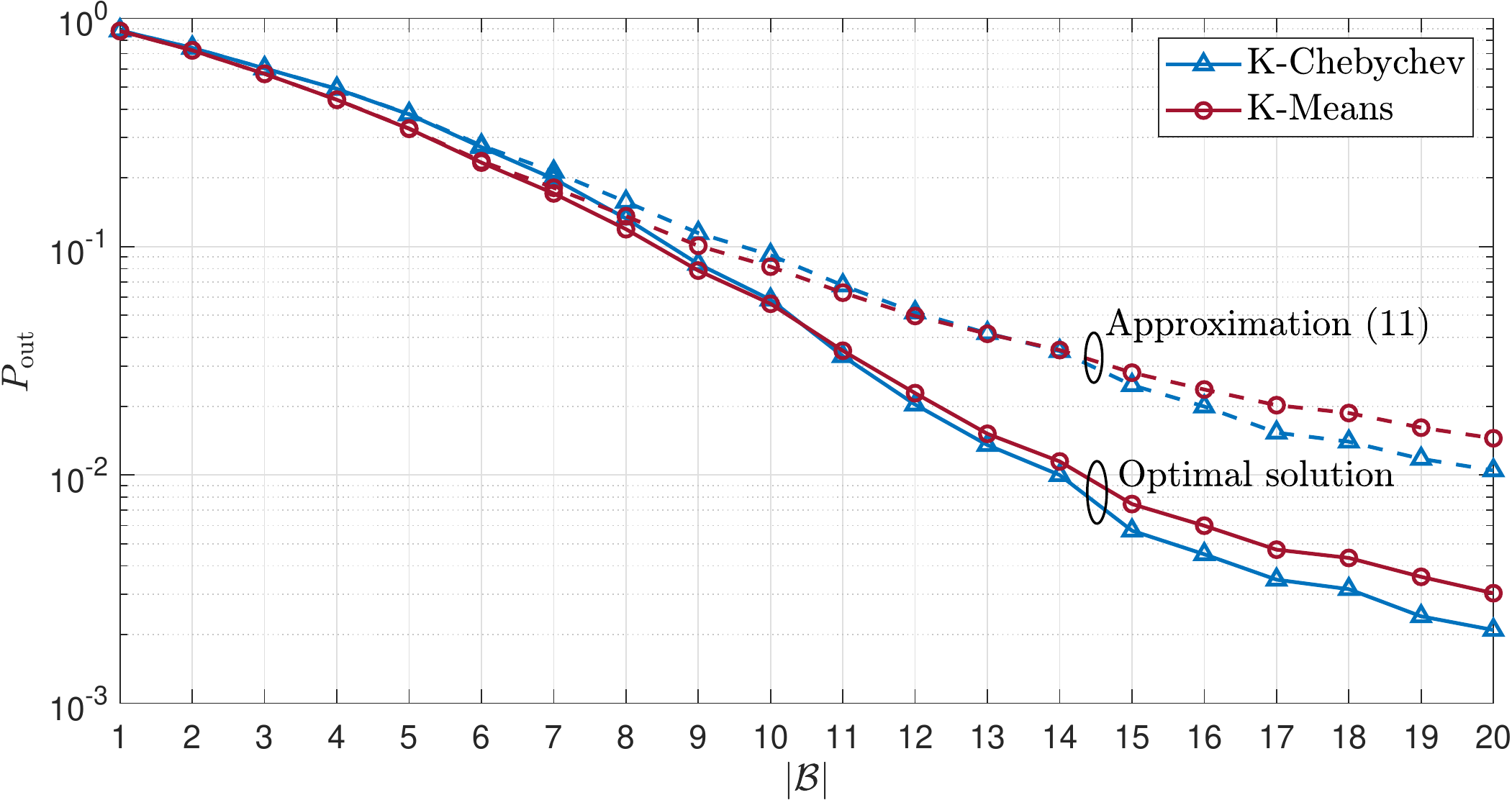}}
	\caption{Network performance vs $|\mathcal{B}|$, for $E_\mathrm{th}= 0.25~J$, in terms of: i) sum transmit power (top) and ii) outage probability (bottom).}
	\vspace*{-2mm}		
	\label{fig:vs_PBs}
\end{figure}

Although the approximation~\eqref{eq:approx} performs poorly for a relatively large number of PBs, we can observe in Figure~\ref{fig:vs_PBs} that when $|\mathcal{B}| \leq 6$ it holds accurate due to the very low contribution of neighbor clusters. The curve $|\mathcal{B}|p_\mathrm{max}$ delimits the feasible region over which the PBs can transmit their energy-carrying signals without violating the constraint \eqref{P1c}.

\section{Summary and Conclusions}\label{section:conclusions}
In this paper, we studied the sum power minimization problem of PBs for powering an IoT deployment. For solving this problem, we first used the devices' positions information to arrange clusters in the network, each headed by a PB, and then optimize the power allocation strategy based on the current devices' battery states. In order to obtain fairer results for the worst-positioned device within each cluster, we proposed a modification of the traditional K-Means algorithm for re-computing the clusters' centroids called K-Chebyshev centering. Numerical results show that this approach achieves better results in terms of outage probability with less sum transmit power when there is a large number of PBs.

\section*{Acknowledgment}
This research has been financially supported by Academy of Finland, 6Genesis Flagship (Grant no318927) and  EE-IoT (no319008), by CNPq, Print CAPES-UFSC ``Automation 4.0'', and RNP/MCTIC (Grant 01245.010604/2020-14) 6G Mobile Communications Systems in Brazil, as well as by ANID FONDECYT Iniciaci\'{o}n No. 11200659 in Chile.

\bibliographystyle{IEEEtran}
\bibliography{IEEEabrv,references}

\end{document}